\documentclass[print,onecolumn,10pt,showpacs,pra]{revtex4}%
\usepackage{amsfonts}
\usepackage{amsmath}
\usepackage{amssymb}
\usepackage{graphicx}%
\begin{document}
\title{Implementation of many-qubit Grover search with trapped ultracold ions }
\author{Wan-Li Yang$^{1,2},$ Hua Wei$^{1,2},$ Chang-Yong Chen$^{3},$ and Mang
Feng$^{1}$}
\email{mangfeng@wipm.ac.cn}
\affiliation{$^{1}$State Key Laboratory of Magnetic Resonance and Atomic and Molecular
Physics, Wuhan Institute of Physics and Mathematics, Chinese Academy of
Sciences, Wuhan 430071, China }
\affiliation{$^{2}$Graduate School of the Chinese Academy of Sciences, Bejing 100049, China}
\affiliation{$^{3}$Department of Physics and Information Engineering, Hunan Institute of
Humanities, Science and Technology, Loudi 417000, China}

\begin{abstract}
We propose a potentially practical scheme for realization of an n-qubit (n%
%TCIMACRO{\TEXTsymbol{>}}%
%BeginExpansion
$>$%
%EndExpansion
2) conditional phase flip (CPF) gate and implementation of Grover search
algorithm in the ion-trap system. We demonstrate both analytically and
numerically that, our scheme could be achieved efficiently to find a marked
state with high fidelity and high success probability. We also show the merits
of the proposal that the increase of the ion number can improve the fidelity
and the success probability of the CPF gate. The required operations for
Grover search are very close to the capabilities of current ion-trap techniques.


\end{abstract}
\maketitle

\section{INTRODUCTION}

Grover's quantum search algorithm [1] has been considered to be an efficient
amplitude-amplification process for quantum states by exploiting the
parallelism of quantum mechanics. As a remarkable idea in quantum computation,
Grover search algorithm could effectively exemplifies the potential speed-up
offered by quantum computers. Recently, many proposals have focused on the
search algorithm by the adiabatic evolution method [2-4] or\ by nonadiabatic
scenario [5]. Although achievement of these schemes needs stringent conditions
and demanding techniques [6], they are really wonderful ideas. On the other
hand, some authors had addressed the effect of decoherence [7], gate
imperfection or errors [8], and noise [9] on the efficiency of quantum
algorithms. We also noticed that, there had been intensive interests in
achieving Grover search algorithm theoretically and experimentally by using
NMR systems [10], linear optical elements [11,12], trapped ions [13-15],
cavity quantum electrodynamics (QED) [16-19], and superconducting mesocircuits [20].

Trapped ions have been considered to be a promising candidate for
quantum-information processing (QIP), due to long coherence time of qubits,
full controllability of operations and high efficiency of detection. In [13],
a many-qubit Grover search algorithm based on the hot-ion quantum computing
[21] in decoherence-free subspace was proposed, where collective dephasing
errors could be kept away from the qubit encoding. Ref. [14] is an alternative
for simulating Grover search algorithm in an ion-trap system, in which the
search could be carried out by more than two qubits. However, in order to
achieve the conditional phase flip (CPF) gate, each ion should be illuminated
by two lasers with different polarizations. In addition, the proposal needs
not only elaborately designed sequences of lasers, but also the auxiliary
states of the ions, which would result in considerable difficulty in
experimental realization with growing number of the qubits.

In this paper, we propose a potentially practical scheme for implementing
n-qubit $(n\geq3)$ Grover search in an ion trap. Compared to above mentioned
schemes, our scheme includes following merits: We carry out the CPF gates with
nearly unity success probability and fidelity in a straightforward way, which
could relax the rigid requirement on accurate sequences of laser manipulations
in previous work, such as sequences for one-qubit and two-qubit gates [12,22]
or swap gates [23]. Moreover, the increase of qubits in our scheme could
improve the fidelity and the success probability of the CPF gate, which are
favorable for a scalable Grover search.

Section II describles the general method for a CPF gate in a linear ion trap.
We then describle, in Sec. IV, the implementation of $n$-qubit Grover search
algorithm based on the proposed CPF gates, which is almost within the reach of
the present technology and extendable to other QIP candidate systems. We
conclude with a discussion in Sec. V.

\section{\textbf{N-QUBIT CPF GATE}}

We consider $n$ identical three-level ions confined in a linear ion trap, as
shown in Fig. 1 where the ions are individually irradiated with traveling wave
laser fields tuned to the first lower vibrational (i.e., red) sideband. The
three-level ionic states under our consideration are denoted by $\left\vert
f_{j}\right\rangle ,\left\vert g_{j}\right\rangle ,$and $\left\vert
e_{j}\right\rangle ,$ with $\left\vert f_{j}\right\rangle $ and $\left\vert
g_{j}\right\rangle $\ being states lower than $\left\vert e_{j}\right\rangle
$. Because the resonant transition happens between $\left\vert g_{j}%
\right\rangle $\ and $\left\vert e_{j}\right\rangle $ by the laser field,
$\left\vert f_{j}\right\rangle $ is not involved in the interaction throughout
our scheme. In the rotating wave approximation, the Hamiltonian in units of
$\hbar=1$ reads,%
\begin{equation}
\ \ \ \allowbreak H=\sum\limits_{j=1}^{n}\{\nu a^{+}a+\omega_{0}\sigma
_{z,j}+[\lambda E^{+}(\mathbf{r},t)\sigma_{j}^{+}+H.c.]\}, \label{1}%
\end{equation}
where $E^{+}(\mathbf{r},t)=E_{0}\exp[-i(kz-\omega_{l}t+\phi)]$ is the positive
frequency part of the driving laser with amplitude $E_{0}$ and frequency
$\omega_{l}=\omega_{0}-\nu.$ $a^{+}$\ $(\allowbreak a)$\ is the creation
(annihilation) operator of the center-of-mass vibrational mode commonly owned
by all the ions, $\sigma_{j}^{+}=\left\vert e_{j}\right\rangle \left\langle
g_{j}\right\vert ,$\ $\sigma_{j}^{-}=\left\vert g_{j}\right\rangle
\left\langle e_{j}\right\vert ,$\ and $\sigma_{z,j}=\frac{1}{2}(\left\vert
e_{j}\right\rangle \left\langle e_{j}\right\vert -\left\vert g_{j}%
\right\rangle \left\langle g_{j}\right\vert )$ are the raising, lowering and
inversion operators for the $j$th trapped ion, respectively. $\nu,$
$\omega_{0}$\ and $\lambda$\ are the trap frequency, the atomic transition
frquency and dipole matrix element (assumed to be real as convention),
respectively. In the resolved sideband limit and in the interaction picture,
the Hamiltonian of Eq. (1) excluding the terms associated with carrier
transition can be simplified to [24],%
\begin{equation}
H_{I}^{^{\prime}}=\sum\limits_{j=1}^{n}\sum\limits_{\varkappa=0}^{\infty
}\Omega_{j}(t)e^{-\eta^{2}/2}e^{i\phi_{j}}\sigma_{j}^{+}\frac{(i\eta
)^{2\varkappa+1}}{\varkappa!(\varkappa+1)!}(a^{+})^{\varkappa}a^{\varkappa
+1}+H.c., \label{2}%
\end{equation}
where $\eta=k/\sqrt{2n\nu M}$ is the Lamb-Dicke parameter with $M$ being the
mass of the ion. We assume the laser pulses to be time-dependent in Gaussian
type, and the time varying Rabi frequency $\Omega_{j}(t)$ of the laser field
is given by%
\begin{equation}
\Omega_{j}(t)=\Omega_{\max}^{j}\exp\{-(t-t_{0})^{2}/2\tau_{j}^{2}\}, \label{3}%
\end{equation}
with $\tau_{j}$ being the duration of the Gaussian shaped pulse irradiating on
the $j$th ion. In the Lamb-Dicke regime, the Hamiltonian can be approximated
by the expansion to the first order in $\eta,$%
\begin{equation}
\ \ \ \allowbreak H_{1}=\sum\limits_{j=1}^{n}i\eta\Omega_{j}(t)(a^{+}%
\sigma_{j}^{-}e^{-i\phi_{j}}-a\sigma_{j}^{+}e^{i\phi_{j}}). \label{4}%
\end{equation}
Furthermore, by assuming $\phi_{j}=\pi/2,$ we reduce Eq. (4) to%
\begin{equation}
\ \allowbreak H_{2}=\sum\limits_{j=1}^{n}\eta\Omega_{j}(t)(a^{+}\sigma_{j}%
^{-}+a\sigma_{j}^{+}). \label{5}%
\end{equation}
\ \ We first assume that the center-of-mass mode of the ions is initially in
the vacuum state $\left\vert 0\right\rangle ,$ and except the last ion (i.e.,
the $n$th ion) initially prepared in the excited state $\left\vert
e_{n}\right\rangle $ and the $k$th ion in the state $\left\vert g_{k}%
\right\rangle ,$ other ions are initially prepared in the state $\left\vert
f\right\rangle $. So we only have the nth and the $k$th ions interacting with
the vibrational mode by following evolution,%
\begin{align}
\prod\nolimits_{j=1,j\neq k}^{n-1}\left\vert g_{k}\right\rangle \left\vert
f_{j}\right\rangle \left\vert e_{n}\right\rangle \left\vert 0\right\rangle  &
\longrightarrow\exp(-i\int H_{2}dt)\prod\limits_{j,j\neq k}^{n-1}\left\vert
g_{k}\right\rangle \left\vert f_{j}\right\rangle \left\vert e_{n}\right\rangle
\left\vert 0\right\rangle \nonumber\\
&  =[\frac{\vartheta_{n}^{2}}{\vartheta_{k}^{^{\prime}2}}\cos(\eta
\vartheta_{k}^{^{\prime}})+\frac{\vartheta_{k}^{^{\prime}2}-\vartheta_{n}^{2}%
}{\vartheta_{k}^{^{\prime}2}}]\times\prod\limits_{j,j\neq k}^{n-1}\left\vert
g_{k}\right\rangle \left\vert f_{j}\right\rangle \left\vert e_{n}\right\rangle
\left\vert 0\right\rangle \nonumber\\
&  +\frac{\vartheta_{n}\vartheta_{k}}{\vartheta_{k}^{^{\prime}2}}[\cos
(\eta\vartheta_{k}^{^{\prime}})-1]\left\vert e_{k}\right\rangle \left\vert
g_{n}\right\rangle \prod\limits_{j,j\neq k}^{n-1}\left\vert f_{j}\right\rangle
\left\vert 0\right\rangle \nonumber\\
&  -\frac{i\vartheta_{n}}{\vartheta_{k}^{^{\prime}}}\sin(\eta\vartheta
_{k}^{^{\prime}})\left\vert g_{k}\right\rangle \left\vert g_{n}\right\rangle
\prod\limits_{j,j\neq k}^{n-1}\left\vert f_{j}\right\rangle \left\vert
1\right\rangle , \label{6}%
\end{align}
where $\vartheta_{k}^{^{\prime}}=\sqrt{\vartheta_{k}^{2}+\vartheta_{n}^{2}},$
and%
\begin{align}
\vartheta_{j}  &  =\int\nolimits_{0}^{2t_{0}}\Omega_{\max}^{j}\exp
\{-(t-t_{0})^{2}/\tau_{j}^{2}\}dt\nonumber\\
&  =\int\nolimits_{-t_{0}}^{t_{0}}\Omega_{\max}^{j}\exp(-t^{2}/2\tau_{j}%
^{2})dt\nonumber\\
&  =\Omega_{\max}^{j}\sqrt{2\pi}\tau_{j}\operatorname{erf}[t_{0}/\sqrt{2}%
\tau_{j}]\nonumber\\
&  \approx\Omega_{\max}^{j}\sqrt{2\pi}\tau_{j}, \label{7}%
\end{align}
where $\operatorname{erf}[z]=(2/\sqrt{\pi})\int\nolimits_{0}^{z}e^{-t^{2}}%
dt$\ is the error function, and $j=1,2,\cdots,n.$ In concrete calculations, we
assume that $\operatorname{erf}[t_{0}/\sqrt{2}\tau_{j}]\longrightarrow1.$
Next, we consider another situation, that is, the last ion is initially in the
excited state $\left\vert e_{n}\right\rangle $, and for other ions some of
which are initially in the state $\left\vert g\right\rangle $ and the rest are
in the state $\left\vert f\right\rangle $, then we can acquire the
corresponding time evolution,%
\begin{align}
\prod\nolimits_{j,l=1,j\neq l}^{n-1}\left\vert g_{j}\right\rangle \left\vert
f_{l}\right\rangle \left\vert e_{n}\right\rangle \left\vert 0\right\rangle  &
\longrightarrow\exp(-i\int H_{2}dt)\prod\limits_{j,l=1,j\neq l}^{n-1}%
\left\vert g_{j}\right\rangle \left\vert f_{l}\right\rangle \left\vert
e_{n}\right\rangle \left\vert 0\right\rangle \nonumber\\
&  =[\frac{\vartheta_{n}^{2}}{\vartheta^{2}}\cos(\eta\vartheta)+\frac
{\vartheta^{2}-\vartheta_{n}^{2}}{\vartheta^{2}}]\times\prod
\limits_{j,l=1,j\neq l}^{n-1}\left\vert g_{j}\right\rangle \left\vert
f_{l}\right\rangle \left\vert e_{n}\right\rangle \left\vert 0\right\rangle
\nonumber\\
&  +\frac{\vartheta_{n}}{\vartheta^{2}}[\cos(\eta\vartheta)-1]\sum
\limits_{k=1}^{s}\vartheta_{k}\left\vert e_{k}\right\rangle \left\vert
g_{n}\right\rangle \prod\limits_{j,l=1,j\neq l}^{n-1}\left\vert g_{j}%
\right\rangle \left\vert f_{l}\right\rangle \left\vert 0\right\rangle
\nonumber\\
&  -\frac{i\vartheta_{j}}{\vartheta}\sin(\eta\vartheta)\prod
\limits_{j,l=1,j\neq l}^{n-1}\left\vert g_{j}\right\rangle \left\vert
f_{l}\right\rangle \left\vert g_{n}\right\rangle \left\vert 1\right\rangle ,
\label{8}%
\end{align}
where $\vartheta=\sqrt{\vartheta_{n}^{2}+\sum\nolimits_{j=1}^{s}\vartheta
_{j}^{2}}$ with $s$ the number of the ions initially in the state $\left\vert
g\right\rangle $ and other denotations are defined as the same as above. To
achieve our aim, we need to consider another case: If the ions are initially
in the state $\prod\nolimits_{j=1}^{n-1}\left\vert f_{j}\right\rangle
\left\vert e_{n}\right\rangle \left\vert 0\right\rangle ,$ then the ions,
except the last one, do not interact with the vibrational mode. The evolution
of the system is as follows,%
\begin{align}
\prod\nolimits_{j=1}^{n-1}\left\vert f_{j}\right\rangle \left\vert
e_{n}\right\rangle \left\vert 0\right\rangle  &  \longrightarrow\exp(-i\int
H_{2}dt)\prod\nolimits_{j=1}^{n-1}\left\vert f_{j}\right\rangle \left\vert
e_{n}\right\rangle \left\vert 0\right\rangle \nonumber\\
&  =[\cos(\eta\vartheta_{n})\left\vert e_{n}\right\rangle \left\vert
0\right\rangle -i\sin(\eta\vartheta_{n})\left\vert g_{n}\right\rangle
\left\vert 1\right\rangle ]\prod\nolimits_{j=1}^{n-1}\left\vert f_{j}%
\right\rangle . \label{9}%
\end{align}
\ \ \ \ Based on Eqs. (6)--(9), we can construct an n-qubit CPF gate. In our
proposal, the qubit definitions are the same for other ions, except the last
ion, i.e., the logic state $\left\vert 1\right\rangle $ ($\left\vert
0\right\rangle $) of the $\allowbreak i$th qubit $\allowbreak$is denoted by
$\left\vert f_{i}\right\rangle $ ($\left\vert g_{i}\right\rangle $) of the
$i$th ion with $i=1,2,\cdots,n-1,$ whereas the logic state $\left\vert
1\right\rangle $ ($\left\vert 0\right\rangle $) of the $n$th qubit is
represented by $\left\vert e_{n}\right\rangle $ ($\left\vert g_{n}%
\right\rangle $) of the $n$th ion. By considering the quantum\ information
encoded in the subspace spanned by the states $\{\left\vert g_{1}\right\rangle
,\left\vert f_{1}\right\rangle ,\left\vert g_{2}\right\rangle ,\left\vert
f_{2}\right\rangle ,\cdots,\left\vert g_{n-1}\right\rangle ,\left\vert
f_{n-1}\right\rangle ,\left\vert g_{n}\right\rangle ,\left\vert e_{n}%
\right\rangle \},$ we have Eq. (9) in the case of $\allowbreak\eta
\vartheta_{n}=\pi$ to be,%
\begin{equation}
\prod\nolimits_{j=1}^{n-1}\left\vert f_{j}\right\rangle \left\vert
e_{n}\right\rangle \left\vert 0\right\rangle \longrightarrow-\prod
\nolimits_{j=1}^{n-1}\left\vert f_{j}\right\rangle \left\vert e_{n}%
\right\rangle \left\vert 0\right\rangle . \label{10}%
\end{equation}
Furthermore,\ we assume that coupling strenghths satisfy the following
condition,%
\begin{equation}
\vartheta_{1}=\vartheta_{2}=\cdots=\vartheta_{n-1}\gg\vartheta_{n}. \label{11}%
\end{equation}
After inserting Eq. (7) into Eq. (11), we have,%
\begin{equation}
m=\frac{\vartheta_{n}}{\vartheta_{i}}=\frac{\Omega_{\max}^{n}\tau_{n}}%
{\Omega_{\max}^{i}\tau_{i}}\ll1, \label{12}%
\end{equation}
with $\allowbreak i=1,2,\cdots,n-1,$ which implies that the condition in Eq.
(11) could be met by adjusting the pulse widths $\tau_{i}$ and the maximum
coupling strength $\Omega_{\max}^{i}.$ To keep the ions in the vacuum
center-of-mass mode, we assume that each laser pulse width has the identical
value ($\tau_{j}=\tau_{0}$), but we have different $\Omega_{\max}^{i}$ by
setting $m=\Omega_{\max}^{n}/\Omega_{\max}^{i}\ll1$ to meet the requirements.
Then Eqs. (6) and (8) can be reduced to
\begin{equation}
\prod\nolimits_{j=1,j\neq k}^{n-1}\left\vert g_{k}\right\rangle \left\vert
f_{j}\right\rangle \left\vert e_{n}\right\rangle \left\vert 0\right\rangle
\longrightarrow\beta\times\prod\limits_{j=1,j\neq k}^{n-1}\left\vert
g_{k}\right\rangle \left\vert f_{j}\right\rangle \left\vert e_{n}\right\rangle
\left\vert 0\right\rangle \label{13}%
\end{equation}
and%
\begin{equation}
\prod\nolimits_{j,l=1,j\neq l}^{n-1}\left\vert g_{j}\right\rangle \left\vert
f_{l}\right\rangle \left\vert e_{n}\right\rangle \left\vert 0\right\rangle
\longrightarrow\alpha_{s}\times\prod\limits_{j,l=1,j\neq l}^{n-1}\left\vert
g_{j}\right\rangle \left\vert f_{l}\right\rangle \left\vert e_{n}\right\rangle
\left\vert 0\right\rangle , \label{14}%
\end{equation}
where $\alpha_{s}=[m^{2}\cos(\sqrt{m^{2}+s}\pi/m)+s]/(m^{2}+s)\ $and
$\beta=[m^{2}\cos(\sqrt{m^{2}+1}\pi/m)+1]/(m^{2}+1)=\alpha_{1}.$ So if the
requirement of Eq. (12) is met,\ we have $\beta\approx\alpha_{s}\approx1$ and
thereby obtain an $n$-qubit CPF gate in our computational subspace, where the
ions from the first to the $(n-1)$th represent the control qubits, and the
last ion represents the target qubit. We take $n=4$ as an example
below$.$\ From Eqs. (10)--(14), an approximate four-qubit CPF gate can be
reached as follows:%
\begin{equation}
U_{CPF}^{(4)}=diag\{1,1,\alpha_{3},\alpha_{2},1,1,\alpha_{2},\beta
,1,1,\alpha_{2},\beta,1,1,\beta,-1\}, \label{15}%
\end{equation}
\ in the computational subspace spanned by \{$\left\vert g_{1}g_{2}g_{3}%
g_{4}\right\rangle ,$ $\left\vert g_{1}g_{2}g_{3}e_{4}\right\rangle ,$
$\left\vert g_{1}g_{2}f_{3}g_{4}\right\rangle ,$ $\left\vert g_{1}g_{2}%
f_{3}e_{4}\right\rangle ,$ $\left\vert g_{1}f_{2}g_{3}g_{4}\right\rangle ,$
$\left\vert g_{1}f_{2}g_{3}e_{4}\right\rangle ,$ $\left\vert g_{1}f_{2}%
f_{3}g_{4}\right\rangle ,$ $\left\vert g_{1}f_{2}f_{3}e_{4}\right\rangle ,$
$\left\vert f_{1}g_{2}g_{3}g_{4}\right\rangle ,$ $\left\vert f_{1}g_{2}%
g_{3}e_{4}\right\rangle ,$ $\left\vert f_{1}g_{2}f_{3}g_{4}\right\rangle ,$
$\left\vert f_{1}g_{2}f_{3}e_{4}\right\rangle ,$ $\left\vert f_{1}f_{2}%
g_{3}g_{4}\right\rangle ,$ $\left\vert f_{1}f_{2}g_{3}e_{4}\right\rangle ,$
$\left\vert f_{1}f_{2}f_{3}g_{4}\right\rangle ,$ $\left\vert f_{1}f_{2}%
f_{3}e_{4}\right\rangle $\}, where $\alpha_{2}=0.99952,$ $\alpha_{3}=0.99516,$
$\beta=0.99988$\ in the case of $m=0.1.$\ 

We now turn to calculation of the fidelity and the success probability
according to the relations $F=\overline{\left\langle \Psi_{0}\right\vert
U_{CPF}^{(n)+}\left\vert \Psi_{f}\right\rangle \left\langle \Psi
_{f}\right\vert U_{CPF}^{(n)}\left\vert \Psi_{0}\right\rangle }$ [25] and
$P=\overline{\langle\Psi_{f}\left\vert \Psi_{f}\right\rangle }$ $,$ where
$\left\vert \Psi_{f}\right\rangle $ is the final state after the $n$-qubit CPF
gating has been made. The overline indicates average over all possible
components in $\left\vert \Psi_{0}\right\rangle .$ In the case of $n$ qubits,
we set $\left\vert \Psi_{0}\right\rangle =\frac{1}{\sqrt{2^{n}}}(\left\vert
g_{1}\right\rangle +\left\vert f_{1}\right\rangle )(\left\vert g_{2}%
\right\rangle +\left\vert f_{2}\right\rangle )\cdots(\left\vert g_{n-1}%
\right\rangle +\left\vert f_{n-1}\right\rangle )(\left\vert g_{n}\right\rangle
+\left\vert e_{n}\right\rangle ),$ the infidelity and the success probability
which gives a general assessment for our gating are obtained straightforwardly
by%
\begin{equation}
Infide\tau ity=1-F=1-\frac{[\sum\nolimits_{s=2}^{n-1}C_{s}^{(n)}\alpha
_{s}+(n-1)\beta+2^{n-1}+1]^{2}}{2^{n}[\sum\nolimits_{s=2}^{n-1}C_{s}%
^{(n)}\alpha_{s}^{2}+(n-1)\beta^{2}+2^{n-1}+1]}, \label{16}%
\end{equation}
and%
\begin{equation}
P=\frac{[\sum\nolimits_{s=2}^{n-1}C_{s}^{(n)}\alpha_{s}^{2}+(n-1)\beta
^{2}+2^{n-1}+1]}{2^{n}}, \label{17}%
\end{equation}
where we denote the number of $\alpha_{s}$ ($\allowbreak s=2,3,\cdots,n-1$) by
$C_{s}^{(n)}$ with $n$ the number of the qubits, and $C_{s}^{(n)}$ fulfilling
the equations,%
\begin{equation}
C_{s}^{(n)}+C_{s+1}^{(n)}=C_{s+1}^{(n+1)}, \label{18}%
\end{equation}
\ and%
\begin{equation}
\sum\nolimits_{s=2}^{n-1}C_{s}^{(n)}=2^{n-1}-n. \label{19}%
\end{equation}
\ We have listed some examples in Table I for the values of $C_{s}^{(n)}$.

\ \ \ \ \ \ \ \ \ \ \ \ \ \ \ \ \ \ \ \ \ \ \ \ \ \ \ \ \ \ \ \ \ \ \ \ TABLE
I. \ List of the values of $C_{s}^{(n)}.$%

\begin{tabular}
[c]{c|c|c|c|c|c|c|c|c}\hline\hline
qubit number $n$ & number of $\alpha_{2}$ & number of $\alpha_{3}$ & number of
$\alpha_{4}$ & number of $\alpha_{5}$ & number of $\alpha_{6}$ & $\cdots$ &
number of $\alpha_{n-1}$ & $\sum\nolimits_{s=2}^{n-1}C_{s}^{(n)}$\\\hline
$n=4$ & 3 & 1 & 0 & 0 & 0 & 0 & 0 & 4\\
$n=5$ & 6 & 4 & 1 & 0 & 0 & 0 & 0 & 11\\
$n=6$ & 10 & 10 & 5 & 1 & 0 & 0 & 0 & 26\\
$n=7$ & 15 & 20 & 15 & 6 & 1 & 0 & 0 & 57\\\hline\hline
\end{tabular}

We have plotted the infidelity$\ $and the success probability versus
$m=\Omega_{\max}^{n}/\Omega_{\max}^{i}$ in Figs. 2 and 3, which clearly
indicate that our proposed $n$-qubit CPF gate is of high fidelity and high
success probability as long as the value of $m$ is small enough. The figures
also show the increase of $F$ and $P$ with the number of the ions. To obtain
the maximal fidelity and success probability, provided that $\Omega_{\max}%
^{i}=$ $\Omega_{m}$,\ we have made some numerical calculations, which show the
values of m to be $m=$ $\Omega_{\max}^{n}/\Omega_{m}=$ 0.0122,\ 0.0122
and\ 0.0147 corresponding to $n=$3, 6 and 9, respectively. One can see from
the figures that the implementation of the $n$-qubit CPF gate with high
fidelity and high success probability can be realized by suitable laser-ion
coupling strength ratio of the last ion to the others.

\section{\textbf{N-QUBIT GROVER SEARCH}}

In this section, we will implement an $n$-qubit Grover algorithm by our gating
discussed above. One of the key steps in Grover search algorithm is to find a
marked element in an unsorted database of size N. Generally speaking, the
Grover search algorithm consists of three kinds of operations [13]. The first
one is to prepare a superposition state $\left\vert \Psi_{0}\right\rangle
=(\frac{1}{\sqrt{2^{n}}})\sum_{i=0}^{2^{n}-1}\left\vert i\right\rangle $ using
Hadamard gates $H^{\otimes n}$ ($n$ $\allowbreak$being the number of qubits).
The second is for an iteration $Q^{(n)}$ including following two operations:
(a) Inverting the amplitude of the marked state $\left\vert \rho\right\rangle
$ using an operator $J_{\rho}=I-2\left\vert \rho\right\rangle \left\langle
\rho\right\vert ,$ with $I$ the identity matrix; (b) Inversion about the
average of the amplitudes of all states using the diffusion transform
$\allowbreak\hat{D}^{(n)}$, with $\hat{D}_{ij}^{(n)}=\tfrac{2}{N}-\delta
_{ij}\ $($i,j=1,2,3,\cdots,\allowbreak N$) and $\allowbreak N=2^{n}$. This
step should be carried out by at least $\pi\sqrt{N}/4$ times to maximize the
probability for finding the marked state. Finally, a measurement of the whole
system is made to get the marked state.

As defined above, the logic state $\left\vert 1\right\rangle $ ($\left\vert
0\right\rangle $) of the $\allowbreak i$th qubit $\allowbreak$is denoted by
$\left\vert f_{i}\right\rangle $ ($\left\vert g_{i}\right\rangle $) of the
$i$th ion, with $i=1,2,\cdots,n-1,$ whereas the logic state $\left\vert
1\right\rangle $ ($\left\vert 0\right\rangle $) of the $n$th qubit is
represented by $\left\vert e_{n}\right\rangle $ ($\left\vert g_{n}%
\right\rangle $) of the $n$th ion. By considering the quantum\ information
encoded in the subspace spanned by $\{\left\vert g_{1}\right\rangle
,\left\vert f_{1}\right\rangle ,\left\vert g_{2}\right\rangle ,\left\vert
f_{2}\right\rangle ,\cdots,\left\vert g_{n-1}\right\rangle ,\left\vert
f_{n-1}\right\rangle ,\left\vert g_{n}\right\rangle ,\left\vert e_{n}%
\right\rangle \},$ we have following transformation,
\begin{equation}
\allowbreak Q^{(n)}=\allowbreak W^{\otimes n}J_{00\cdots0}^{(n)}\allowbreak
W^{\otimes n}J_{\rho}=\allowbreak W^{\otimes n}\allowbreak J_{g_{1}g_{2}\cdots
g_{n}}^{(n)}W^{\otimes n}J_{\rho}=-\hat{D}^{(n)}J_{\rho}, \label{20}%
\end{equation}
where $J_{00\cdots0}^{(n)}=diag\{-1,1,\cdots,1\}=I^{(n)}-2\left\vert
00\cdots0\right\rangle \left\langle 00\cdots0\right\vert $ (the number of
$0$\ is $n$) and the Hadamard gate in our whole computational subspace is
given by%

\begin{equation}
W^{\otimes n}=\prod\limits_{i=1}^{n}W_{i}=\left(  \frac{1}{\sqrt{2}}\right)
^{n}%
\begin{bmatrix}
1 & 1\\
1 & -1
\end{bmatrix}
\otimes%
\begin{bmatrix}
1 & 1\\
1 & -1
\end{bmatrix}
\otimes\cdots\otimes%
\begin{bmatrix}
1 & 1\\
1 & -1
\end{bmatrix}
. \label{21}%
\end{equation}
In the case of $n$ qubits, Eq. (20) implies that the diffusion transform
$\hat{D}^{(n)}=-\allowbreak W^{\otimes n}\allowbreak J_{g_{1}g_{2}\cdots
g_{n}}^{(n)}W^{\otimes n}$ is always unchanged no matter which state is to be
searched. The only change is the operator $J_{\rho}$ for different marked
states. Based on the CPF gate constructed in last section, we have
$J_{11\cdots1}^{(n)}\approx diag\{1,1,\cdots,-1\}=U_{CPF}^{(4)},$ from which
we could achieve the gate $J_{00\cdots0}^{(n)}$ and other relevant operations,%
\begin{equation}
J_{00\cdots0}^{(n)}=\sigma_{x,n}S_{x,n-1}\cdots S_{x,1}J_{11\cdots1}%
^{(n)}S_{x,1}\cdots S_{x,n-1}\sigma_{x,n}, \label{22}%
\end{equation}
and%
\begin{equation}
\hat{D}^{(n)}=\allowbreak W^{\otimes n}J_{00\cdots0}^{(n)}\allowbreak
W^{\otimes n}, \label{23}%
\end{equation}
where $S_{x,j}=|f_{j}\rangle\langle g_{j}|+|g_{j}\rangle\langle f_{j}|$ with
$j\neq n,$ and $\sigma_{x,n}=|e_{n}\rangle\langle g_{n}|+|g_{n}\rangle\langle
e_{n}|.$ To achieve $\allowbreak Q^{(n)}$, we will construct the CPF gate
$J_{\rho}=I-2\left\vert \rho\right\rangle \left\langle \rho\right\vert .$ For
example, in the case of $n=4$, the number of possible quantum states is
$\allowbreak2^{4}$, and the operation is to label a marked state by
$\allowbreak J_{\rho}$ with $\rho$ one of the states from $\{\left\vert
0000\right\rangle ,\left\vert 0001\right\rangle ,\left\vert 0010\right\rangle
,\cdots,\left\vert 1111\right\rangle \}.$ To carry out the four-qubit Grover
search, we need two four-qubit Hadamard gates $W^{\otimes4}$. Based on the
gate $J_{1111}=U_{CPF}^{(4)}$ (See Eq. (15)) marking the state $\left\vert
f_{1}f_{2}f_{3}e_{4}\right\rangle ,$ we could construct other fifteen gates
for the marking job as,%
\begin{align}
J_{g_{1}g_{2}g_{3}g_{4}}  &  =J_{0000}=\sigma_{x,4}S_{x,3}S_{x,2}%
S_{x,1}J_{1111}S_{x,1}S_{x,2}S_{x,3}\sigma_{x,4},\ J_{g_{1}g_{2}g_{3}e_{4}%
}=J_{0001}=S_{x,3}S_{x,2}S_{x,1}J_{1111}S_{x,1}S_{x,2}S_{x,3},\ \ \nonumber\\
J_{g_{1}g_{2}f_{3}g_{4}}  &  =J_{0010}=\sigma_{x,4}S_{x,2}S_{x,1}%
J_{1111}S_{x,1}S_{x,2}\sigma_{x,4},\text{ \ }J_{g_{1}g_{2}f_{3}e_{4}}%
=J_{0011}=S_{x,2}S_{x,1}J_{1111}S_{x,1}S_{x,2},\ \nonumber\\
\ J_{g_{1}f_{2}g_{3}g_{4}}  &  =J_{0100}=\sigma_{x,4}S_{x,3}S_{x,1}%
J_{1111}S_{x,1}S_{x,3}\sigma_{x,4},\ \ \ J_{g_{1}f_{2}g_{3}e_{4}}%
=J_{0101}=S_{x,3}S_{x,1}J_{1111}S_{x,1}S_{x,3},\nonumber\\
\text{ }J_{g_{1}f_{2}f_{3}g_{4}}  &  =J_{0110}=\sigma_{x,4}S_{x,1}%
J_{1111}S_{x,1}\sigma_{x,4},\ \ J_{g_{1}f_{2}f_{3}e_{4}}=J_{0111}%
=S_{x,1}J_{1111}S_{x,1}\nonumber\\
J_{f_{1}g_{2}g_{3}g_{4}}  &  =J_{1000}=\sigma_{x,4}S_{x,3}S_{x,2}%
J_{1111}S_{x,2}S_{x,3}\sigma_{x,4},\ J_{f_{1}g_{2}g_{3}e_{4}}=J_{1001}%
=S_{x,3}S_{x,2}J_{1111}S_{x,2}S_{x,3},\nonumber\\
J_{f_{1}g_{2}f_{3}g_{4}}  &  =J_{1010}=\sigma_{x,4}S_{x,2}J_{1111}%
S_{x,2}\sigma_{x,4},\ J_{f_{1}g_{2}f_{3}e_{4}}=J_{1011}=S_{x,2}J_{1111}%
S_{x,2},\nonumber\\
J_{f_{1}f_{2}g_{3}g_{4}}  &  =J_{1100}=\sigma_{x,4}S_{x,3}J_{1111}%
S_{x,3}\sigma_{x,4},\ J_{f_{1}f_{2}g_{3}e_{4}}=J_{1101}=S_{x,3}J_{1111}%
S_{x,3},\ J_{f_{1}f_{2}f_{3}g_{4}}=J_{1110}=\sigma_{x,4}J_{1111}\sigma_{x,4}.
\label{24}%
\end{align}
So with a state marked, and the four-qubit diffusion transform $\hat{D}^{(4)}%
$\ which is generated by combining two Hadamard gates $W^{\otimes4}$ with
$J_{0000}$, a full Grover search for four qubits is available. In principle,
if each component of the design is available, our scheme would be achievable experimentally.

Taking the marked state $\left\vert f_{1}g_{2}g_{3}e_{4}\right\rangle $ as an
example, a standard quantum circuit for the Grover search algorithm for $N=4$
entries is shown in Fig. 4. The procedure of the Grover search is accomplished
with three primitive and sequential steps. First we should perform phase
inversion of the desired state by realizing$\allowbreak J_{\rho}=J_{f_{1}%
g_{2}g_{3}e_{4}}=S_{x,3}S_{x,2}J_{1111}^{(4)}S_{x,2}S_{x,3}.$ Then we invert
all the states with respect to the average amplitude of all the states, which
can be achieved by making use of diffusion transform $\hat{D}^{(4)}%
=\allowbreak W^{\otimes4}J_{0000}^{(4)}W^{\otimes4}=W^{\otimes4}\sigma
_{x,4}S_{x,3}S_{x,2}S_{x,1}J_{1111}^{(4)}S_{x,1}S_{x,2}S_{x,3}\sigma
_{x,4}W^{\otimes4}.$ Finally, a measurement of the whole system is made to
obtain the marked state $\left\vert f_{1}g_{2}g_{3}e_{4}\right\rangle $. By
repeating the first two steps for several times, we can get the marked state
with high success probability. However, to simplify above considerations, we
can reduce the indispensable single-qubit rotations (i.e., $W,$ $S_{x},$ and
$\sigma_{x}$) to the corresponding transforms $R_{i}$ and $R_{i}^{^{\prime}}$
($i=$ 1, 2 , 3, 4),%
\begin{equation}
R_{1}=R_{4}=\left[
\begin{array}
[c]{cc}%
1 & 1\\
-1 & 1
\end{array}
\right]  ,\ \ R_{2}=R_{3}=\left[
\begin{array}
[c]{cc}%
-1 & 1\\
1 & 1
\end{array}
\right]  ,\ \ R_{1}^{^{\prime}}=R_{2}^{^{\prime}}=R_{3}^{^{\prime}}%
=R_{4}^{^{\prime}}=\left[
\begin{array}
[c]{cc}%
1 & -1\\
1 & 1
\end{array}
\right]  .\ \label{25}%
\end{equation}
\ As the qubits 1, 2, and 3 are encoded in the Zeeman sublevels, the
transforms $R_{i}$ and $R_{i}^{^{\prime}}$ ($i=$ 1, 2 , 3) should be carried
out by a pair of laser beams in Raman process [26], whereas $R_{4}$ or
$R_{4}^{^{\prime}}$ operating on the 4th ion could be achieved by using only
one laser. Along with $J_{1111}^{(4)}$ described in above section, all the
necessary operations for Grover search are available in a straightforward way.
Therefore, based on the reliable control of the laser pulses on stationary
ions, we could carry out a four-qubit Grover algorithm with four-qubit CPF
gates and a series of single-qubit gatings according to the operations plotted
in Fig. 4. As a result, we consider that our scheme could significantly reduce
the overhead for QIP tasks with trapped ions due to direct implementation of
n-qubit CPF gates.

As the Grover search involving more than two qubits is carried out only
probabilistically, how to obtain an optimal search is a problem of much
interest. Considering the success probability of the CPF gate (i.e., Eq.
(17))\ and the intrinsic probability of the Grover search itself, we show by
numerical simulation the implementation of the Grover search in Fig. 5. Due to
the similarity, we only demonstrate the search for a marked state in the case
of $n=$ $\allowbreak4$ and $\allowbreak5,$ which demonstrates the decreasing
success rates with the increase of $m$ and the qubit number.

\section{\textbf{DISCUSSION AND CONCLUSION}}

We briefly address the experimental feasibility of our scheme. We may employ
$S_{1/2}(m_{j}=1/2),$ $S_{1/2}(m_{j}=-1/2),$ and $D_{5/2}(m_{j}=-1/2)$
of$\ ^{40}\allowbreak Ca^{+}$\ [27] as the states $\left\vert f\right\rangle
,$ $\left\vert g\right\rangle $ and $\left\vert e\right\rangle ,$
respectively. The lifetime of $D_{5/2}$\ of the $^{40}\allowbreak Ca^{+}$ ion
is longer than $\allowbreak1\allowbreak$ $\sec,$ and the transition between
$\left\vert f\right\rangle $ and $\left\vert g\right\rangle $ is dipolar
forbidden$.$ To carry out the CPF gate in the present scheme, the
condition\emph{ }$\eta\Omega_{\max}^{n}\sqrt{2\pi}\tau_{n}=\pi$\emph{ }and the
approximation relation $\operatorname{erf}[t_{0}/\sqrt{2}\tau_{j}%
]\longrightarrow1$ should be well satisfied. By choosing suitable values of
the ($z$-axis) trap frequency $\nu$ and maximal Rabi frequency $\Omega_{m}$ in
current ion trap experiments, all the above-mentioned relations could be met
in the weak-excitation regime $\Omega_{m}\ll\eta\nu$ (which implies a
negligible AC Stark shift induced by the radiation), where one may expect the
unwanted off-resonant transition to be disregarded on the precondition that
the laser field cannot be made too intense at any rate, which thereby is
regarded as the dominating limiting factor on the resulting relatively-low
gate speed [28]. Taking our proposed $n$-qubit CPF gate as an example, for a
typical value $\eta=0.1$ of the Lamb-Dicke parameter, we have listed the
required time $t_{0}$ for CPF gate operation in Table II, which clearly
indicate that the values of $t_{0}$ are prolonged with the growth of the
number of ions. In addition, the switching rate of the CPF gate will be of
order $\lesssim\nu/1000,$ which, from the viewpoint of heating effect, will
inevitably bring out disadvantaged impact on the construction of CPF gate.
Alternatively, in what follows we will take into account another different
regimes $\Omega_{m}\ll\nu/\eta,$ where the only new requirement we should make
is that the laser should take a fixed intensity satisfaying the resonance
condition [28] $\Omega_{m}=\nu/2,$ which implies an improvement by two orders
of magnitude with respect to the CPF gating time in the case of
weak-excitation. So to reduce the implementation time $t_{0}$, we may choose
such a stronger radiation with $\Omega_{m}=\nu/2$ for a carrier transition.
Our direct calculation based on the model in [28] shows that we could also
have Eq. (5) due to AC Stark shift and we could should satisfy $t_{0}%
\geqslant0.574ms,$ $0.766ms,$ $0.957ms$ in the case of $n=3,$ $4,$ $5$,
respectively, to achieve our CPF gate. As the heating time of the ground
vibrational state of the ions in the linear trap is about\emph{ 4 millisec
[29]}, the implementation time in the case of strong radiation seems better in
view of avoiding heating. Note also that single-qubit operation takes
negligible time in comparison with that for many-qubit phase gating, so direct
calculation shows that one iteration of our proposed Grover search would take
$2t_{0}\approx1ms,$ which is shorter than the the heating time of the ions. In
this sense, the decoherence originating from the heating effect is not a big
obstacle for the simulation of Grover search by our proposal when the number
of qubit is small.

\ \ \ \ \ \ \ \ \ \ \ \ \ TABLE II. \ List of the required time $t_{0}$ for
CPF gate and the values of $\nu$ and $\Omega_{m}$ ($\eta=0.1$).

\ \ \ \ \ \ \ \ \ \ \ \ \ \ \
\begin{tabular}
[c]{c|c|c|c|c|c|c|c|c}\hline
regime & qubit number $n$ & $n=2$ & $n=3$ & $n=4$ & $n=5$ & $n=6$ & $n=7$ &
$n=8$\\\hline
& \ $\nu$ $\ (MHz/2\pi)$ & 0.250 & 0.167 & 0.125 & 0.100 & 0.083 & 0.071 &
0.063\\\hline
$\Omega_{m}\ll\eta\nu$ & $\ \Omega_{m}$ $\ (kHz/2\pi)$ & 2.500 & 1.670 &
1.250 & 1.000 & 0.830 & 0.710 & 0.630\\\hline
& $t_{0}$ $\ \ (ms)$ & 19.14 & 28.71 & 38.29 & 47.86 & 57.43 & 67.00 &
76.57\\\hline
$\Omega_{m}\ll\nu/\eta$ & $\Omega_{m}$ $\ (MHz/2\pi)$ & 0.125 & 0.084 &
0.063 & 0.050 & 0.042 & 0.036 & 0.032\\\hline
& $t_{0}$ $\ \ (ms)$ & 0.383 & 0.574 & 0.766 & 0.957 & 1.148 & 1.340 &
1.531\\\hline
\end{tabular}
\ \ \ \ \ \ \ 

It is generally considered that the computing operation on trapped ions in a
linear ion trap would be more and more intricate with the increase of the
number of ions. Because of the decrease of the spatial separation of the
trapped ions, individual manipulation is more and more difficult, and
meanwhile the vibrational mode spectrum becomes more and more unresolved. As a
result, the extension of quantum computation from a few qubits to large
numbers of qubits is quite technically challenging. In this sense, although
our proposal is in principle scalable, the currently technical level regarding
the linear trap restricts the application of our scheme. Nevertheless, eight
ultracold ions in the linear trap could be individually addressed and
entangled [30]. Therefore, if we apply our scheme to these eight ions, a
Gorver search with $2^{8}$ states could be achieved in a simpler way.

Alternatively, we may consider the application of our scheme in a multi-zone
trap, in which we may carry out our scheme on few ions in separate zones,
respectively, and then entangle a large numbers of ions by moving the ions
between different zones [31]. This is a possible way to a large-scale Grover
search implementation.

We have reiterated our scheme to achieve a CPF gate in one step, which could
save the implementation time and steps compared to conventional methods.
Moreover the most impressive feature of our scheme is the higher fidelity and
higher success rate with more qubits involved, which favors large-scale QIP.
Nevertheless, the more qubits, the more steps necessary for an optimal Grover
search. As the CPF gate proposed here is intrinsically imperfect, a decrease
of the fidelity is inevitable with more qubits involved, even if we neglect
the unpredictable imperfection and decoherence in actually experimental
situation. Anyway, the curent experimental progress has shown some efficient
ways to deal with the imperfection [32]. Other basic operations are also
available because the qubit encoding in the first n-1 ions is like that done
in Oxford's group [26] and the qubit encoding in the last ion is employed in
Innsbruck's group [27]. So we believe that our proposal for Grover search
should be available with current or near-future techniques.

In summary, a potentially practical scheme for performing $n$-qubit CPF gate
as well as the many-qubit Grover search algorithm has been proposed in the
ion-trap system. We have demonstrated both analytically and numerically that,
our scheme could be achieved efficiently to find a marked state with high
fidelity and high success probability. Our proposal could be employed in both
the linear ion trap and the multizone trap, and may also be applied to other
QIP tasks, such as preparation of cluster states for one-way quantum
computation [33]. Therefore, we argue that our idea in the present paper is
not only practical, but also simple and experimentally feasible, which would
be helpful for working in large-scale QIP devices.

\section{\textbf{ACKNOWLEDGMENTS}}

We would like to thank Z. J. Deng and F. Zhou for enlightening discussions.
This work is partly supported by NNSF of China under Grants No. 10774163 and
No. 10774042, and partly by the NFRP of China under Grants No. 2005CB724502
and No. 2006CB921203.

\bigskip

[1] L. K. Grover, \textquotedblleft Quantum mechanics helps in searching for a
needle in a haystack,\textquotedblright\ Phys. Rev. Lett. \textbf{79}, 325 (1997).

[2] D. Daems and S. Gu\'{e}rin, \textquotedblleft Adiabatic quantum search
scheme with atoms in a cavity driven by lasers,\textquotedblright\ Phys. Rev.
Lett. \textbf{99}, 170503 (2007).

[3] J. Roland and N. J. Cerf, \textquotedblleft Quantum search by local
adiabatic evolution,\textquotedblright\ Phys. Rev. A \textbf{65}, 042308 (2002).

[4] R. Sch\"{u}tzhold and G. Schaller, \textquotedblleft Adiabatic quantum
algorithms as quantum phase transitions: First versus second
order,\textquotedblright\ Rev. A \textbf{74}, 060304(R) (2006); M. Stewart
Siu, \textquotedblleft Adiabatic rotation, quantum search, and preparation of
superposition states,\textquotedblright\ Phys. Rev. A \textbf{75}, 062337 (2007).

[5] A. P\'{e}rez and A. Romanelli, \textquotedblleft Nonadiabatic quantum
search algorithms,\textquotedblright\ Phys. Rev. A \textbf{76}, 052318 (2007).

[6] D. M. Tong, K. Singh, L. C. Kwek, and C. H. Oh, \textquotedblleft
Sufficiency criterion for the validity of the adiabatic
approximation,\textquotedblright\ Phys. Rev. Lett. \textbf{98,} 150402 (2007);
Zhaohui Wei and Mingsheng Ying, \textquotedblleft Quantum adiabatic
computation and adiabatic conditions,\textquotedblright\ Phys. Rev. A
\textbf{76}, 024304 (2007).

[7] Hiroo Azuma, \textquotedblleft Higher-order perturbation theory for
decoherence in Grover's algorithm,\textquotedblright\ Phys. Rev. A
\textbf{72}, 042305 (2005); M. Tiersch and R. Sch\"{u}tzhold,
\textquotedblleft Non-Markovian decoherence in the adiabatic quantum search
algorithm,\textquotedblright\ Phys. Rev. A \textbf{75}, 062313 (2007).

[8] Gui Lu Long, Yan Song Li, Wei Lin Zhang, and Chang Cun Tu,
\textquotedblleft Dominant gate imperfection in Grover's quantum search
algorithm,\textquotedblright\ Phys. Rev. A \textbf{61}, 042305 (2000).

[9] B. Pablo-Norman and M. Ruiz-Altaba, \textquotedblleft Noise in Grover's
quantum search algorithm,\textquotedblright\ Phys. Rev. A \textbf{61}, 012301
(1999); D. Shapira, S. Mozes, and O. Biham, \textquotedblleft\ The effect of
unitary noise on Grover's quantum search algorithm,\textquotedblright\ eprint
quant-ph/0307142; P. J. Salas, \textquotedblleft Noise effect on Grover
algorithm,\textquotedblright\ eprint quant-ph/0801.1261.

[10] J. A. Jones, M. Mosca, R. H. Hansen, \textquotedblleft Implementation of
a quantum search algorithm on a quantum computer,\textquotedblright\ Nature
(London) \textbf{393}, 344-346 (1998); I. L. Chuang, N. Gershenfeld, and M.
Kubinec, \textquotedblleft Experimental implementation of fast quantum
searching,\textquotedblright\ Phys. Rev. Lett. \textbf{80}, 3408 (1998).

[11] P. Walther, K. J. Resch, T. Rudolph, E. Schenck, H. Weinfurter, V.
Vedral, M. Aspelmeyer, and A. Zeilinger, \textquotedblleft Experimental
one-way quantum computing,\textquotedblright\ Nature (London) \textbf{434},
169-176 (2005); P. G. Kwiat, J. R. Mitchell, P. D. D. Schwindt, and A. G.
White, \textquotedblleft Grover's search algorithm: An optical
approach,\textquotedblright\ J. Mod. Opt. \textbf{47}, 257 (2000).

[12] J. L. Dodd, T. C. Ralph, and G. J. Milburn, \textquotedblleft
Experimental requirements for Grover's algorithm in optical quantum
computation,\textquotedblright\ Phys. Rev. A \textbf{68}, 042328 (2003).

[13] M. Feng, \textquotedblleft Grover search with pairs of trapped
ions,\textquotedblright\ Phys. Rev. A \textbf{63}, 052308 (2001).

[14] S. Fujiwara and S. Hasegawa, \textquotedblleft General method for
realizing the conditional phase-shift gate and a simulation of Grover's
algorithm in an ion-trap system,\textquotedblright\ Phys. Rev. A \textbf{71},
012337 (2005).

[15] K.-A. Brickman, P. C. Haljan, P. J. Lee, M. Acton, L. Deslauriers, and C.
Monroe, \textquotedblleft Implementation of Grover's quantum search algorithm
in a scalable system,\textquotedblright\ Phys. Rev. A \textbf{72}, 050306(R) (2005).

[16] A. Rauschenbeutel, G. Nogues, S. Osnaghi, P. Bertet, M. Brune, J. M.
Raimond, and S. Haroche, \textquotedblleft Coherent operation of a tunable
quantum phase gate in cavity QED,\textquotedblright\ Phys. Rev. Lett.
\textbf{83}, 5166 (1999); S. Osnaghi, P. Bertet, A. Auffeves, P. Maioli, M.
Brune, J. M. Raimond, and S. Haroche, \textquotedblleft\textquotedblright%
\ Phys. Rev. Lett. \textbf{87}, 037902 (2001).

[17] F. Yamaguchi, P. Milman, M. Brune, J. M. Raimond, and S. Haroche,
\textquotedblleft Quantum search with two-atom collisions in cavity
QED,\textquotedblright\ Phys. Rev. A \textbf{66}, 010302(R) (2002).

[18] W. L. Yang, C. Y. Chen, and M. Feng, \textquotedblleft Implementation of
three-qubit Grover search in cavity quantum electrodynamics,\textquotedblright%
\ Phys. Rev. A \textbf{76}, 054301 (2007); Z. J. Deng, M. Feng, and K. L. Gao,
\textquotedblleft Simple scheme for the two-qubit Grover search in cavity
QED,\textquotedblright\ Phys. Rev. A \textbf{72}, 034306 (2005).

[19] A. Joshi and M. Xiao, \textquotedblleft Three-qubit quantum-gate
operation in a cavity QED system,\textquotedblright\ Phys. Rev. A \textbf{74},
052318 (2006).

[20] Y. Nakamura, Yu. A. Pashkin, and J. S. Tsai, \textquotedblleft Coherent
control of macroscopic quantum states in a single-Cooper-pair
box,\textquotedblright\ Nature (London) \textbf{398}, 786-788 (1999); D. Vion,
A. Aassime, A. Cottet, \textquotedblleft Manipulating the Quantum State of an
Electrical Circuit,\textquotedblright\ Science \textbf{296}, 886-889 (2002);
M. S. Anwar, D. Blazina, H. A. Carteret, S. B. Duckett, J. A. Jones,
\textquotedblleft Implementing Grover's quantum search on a para-hydrogen
based pure state NMR quantum computer,\textquotedblright\ Chem. Phys. Lett.
\textbf{400}, 94-97 (2004).

[21] K. M\o lmer and A. S\o rensen, \textquotedblleft Multiparticle
entanglement of hot trapped ions,\textquotedblright\ Phys. Rev. Lett.
\textbf{82}, 1835 (1999); A. S\o rensen and K. M\o lmer, \textquotedblleft
Quantum computation with ions in thermal motion,\textquotedblright%
\ \textit{ibid.} \textbf{82}, 1971 (1999).

[22] M. J. Bremner, C.M. Dawson, J. L. Dodd, A. Gilchrist, A.W. Harrow, D.
Mortimer, M. A. Nielsen, and T. J. Osborne, \textquotedblleft Practical scheme
for quantum computation with any two-qubit entangling gate,\textquotedblright%
\ Phys. Rev. Lett. \textbf{89}, 247902 (2002); S. Lloyd, \textquotedblleft
Almost any quantum logic gate is universal,\textquotedblright\ Phys. Rev.
Lett. \textbf{75}, 346 (1995).

[23] N. Schuch and J. Siewert, \textquotedblleft Programmable networks for
quantum algorithms,\textquotedblright\ Phys. Rev. Lett. \textbf{91}, 027902 (2003).

[24] For example, R. L. de Matos Filho and W. Vogel, \textquotedblleft Even
and odd coherent states of the motion of a trapped ion,\textquotedblright%
\ Phys. Rev. Lett. \textbf{76,} 608 (1996); W. Vogel\ and R. L. de Matos
Filho, \textquotedblleft Nonlinear Jaynes-Cummings dynamics of a trapped
ion,\textquotedblright\ Phys. Rev. A \textbf{52}, 4214 (1995).\ For clarity of
description below, we first assume a weak radiation without inducing AC Stark
shift. So we consider in Eq. (2) the resonant excitation involving motional
quanta exchange. Actually, with a strong radiation, we could also do our job
by carrier transition, as discussed later.

[25] J. F. Poyatos, J. I. Cirac, and P. Zoller, \textquotedblleft Complete
characterization of a quantum process: The two-bit quantum
gate,\textquotedblright\ Phys. Rev. Lett. \textbf{78,} 390 (1997).

[26] M. J. McDonnell, J.-P. Stacey, S. C. Webster, J. P. Home, A. Ramos, D. M.
Lucas, D. N. Stacey, and A. M. Steane, \textquotedblleft High-efficiency
detection of a single quantum of angular momentum by suppression of optical
pumping,\textquotedblright\ Phys. Rev. Lett. \textbf{93}, 153601 (2004).

[27] C. F. Roos, M. Riebe, H. Haffner, W. Hansel, J. Benhelm, G. P. T.
Lancaster, C. Becher,\ F. Schmidt-Kaler, and R. Blatt, \textquotedblleft
Control and Measurement of Three-Qubit Entangled States,\textquotedblright%
\ Science \textbf{304}, 1478-1480 (2004).

[28] D. Jonathan, M. B. Plenio and P. L. Knight, \textquotedblleft Fast
quantum gates for cold trapped ions,\textquotedblright\ Phys. Rev. A\textbf{
62}, 042307\ (2000).

[29] M.D. Barrett, J. Chiaverini, T. Schaetz, J. Britton, W.M. Itano, J.D.
Jost, E. Knill, C. Langer, D. Leibfried, R. Ozeri, D.J. Wineland,
\textquotedblleft Deterministic quantum teleportation of atomic
qubits,\textquotedblright\ Nature \textbf{429}, 737-739 (2004).

[30] H. H\"{a}ffner, W. H\"{a}nsel, C. F. Roos, J. Benhelm, D. Chek-alkar, M.
Chwalla, T. K\"{o}rber, U. D. Rapol, M. Riebe, P. O. Schmidt, C. Becher, O.
G\"{u}hne, W. D\"{u}r, and R. Blatt, \textquotedblleft Scalable multiparticle
entanglement of trapped ions,\textquotedblright\ Nature (London) \textbf{438},
643-646 (2005).

[31] D. Kielpinski, C. Monroe, and D. J. Wineland, \textquotedblleft
Architecture for a large-scale ion-trap quantum computer,\textquotedblright%
\ Nature (London)\textbf{ 417}, 709-711 (2002); D. Leibfried, E. Knill, C.
Ospelkaus, and D. J. Wineland, \textquotedblleft Transport quantum logic gates
for trapped ions,\textquotedblright\ Phys. Rev. A. \textbf{76}, 032324 (2007).

[32] D. Leibfried, B. DeMarco, V. Meyer, M. Rowe, A. Ben-Kish, M. Barrett, J.
Britton, J. Hughes, W. M. Itano, B. M. Jelenkovic, C. Langer, D. Lucas, T.
Rosenband, and D. J. Wineland, \textquotedblleft Towards quantum information
with trapped ions at NIST,\textquotedblright\ J. Phys. B\textbf{ 36}, 599 (2003).

[33] R. Raussendorf and H. J. Briegel, \textquotedblleft A one-way quantum
computer,\textquotedblright\ Phys. Rev. Lett. \textbf{86,} 5188 (2001).

\textbf{Captions of Figures}

FIG. 1. Schematic setup for implementing n-qubit CPF gate and Grover search in
a linear trap, where the inset shows the ionic level configuration, with bold
lines for the states encoding qubits and the arrows for the coupling of the
lasers to the ions.

FIG. 2. Infidelity versus the Rabi frequency ratio of the last ion to the
others, where the solid, dashed-dotted and dashed curves represent the case of
$n=3,$ $6$ and $9$, respectively.

FIG. 3. Success probability versus the Rabi frequency ratio of the last ion to
the others, where the solid, dashed-dotted and dashed curves represent the
case of $n=3,$ $6$ and $9$, respectively.

FIG. 4. Quantum circuit of one iteration of the four-qubit Grover search for
the marked state $\left\vert f_{1}g_{2}g_{3}e_{4}\right\rangle $, where $W,$
$J_{1111}^{(4)}$ and $S_{x}$ ($\sigma_{x}$) are the Hadamard gate, four-qubit
controlled phase gate and single-qubit NOT gate, respectively. The state of
ions is initially prepared in the average superposition state $\left\vert
\Psi_{I}\right\rangle =\frac{1}{4}(\left\vert g_{1}\right\rangle +\left\vert
f_{1}\right\rangle )(\left\vert g_{2}\right\rangle +\left\vert f_{2}%
\right\rangle )(\left\vert g_{3}\right\rangle +\left\vert f_{3}\right\rangle
)(\left\vert g_{4}\right\rangle +\left\vert e_{4}\right\rangle ).$ The
operations in the dashed boxes could be reduced to the transforms $R_{i}$ and
$R_{i}^{^{\prime}}$ with $i($=1, 2, 3, 4) denoting the $ith$ ion, which is
helpful for experimental implementation. To maximize the search propability,
we should implement the circuit repeatedly for several times.

FIG. 5. The search probability for the marked states $\left\vert f_{1}%
g_{2}g_{3}e_{4}\right\rangle $ and $\left\vert f_{1}g_{2}g_{3}e_{4}%
e_{5}\right\rangle $ versus the number of the iterations in the case of $n=4$
and $5.$

\newpage%
%TCIMACRO{\FRAME{ftbpF}{5.617in}{3.6115in}{0in}{}{}{1.eps}%
%width 5.617in;  height 3.6115in;  depth 0in;  original-width 5.559in;
%original-height 3.5639in;  cropleft "0";  croptop "1";  cropright "1";
%cropbottom "0";  filename '1.eps';file-properties "XNPEU";}}}%
%BeginExpansion
\begin{figure}
[ptb]
\begin{center}
\includegraphics[
height=3.6115in,
width=5.617in
]%
{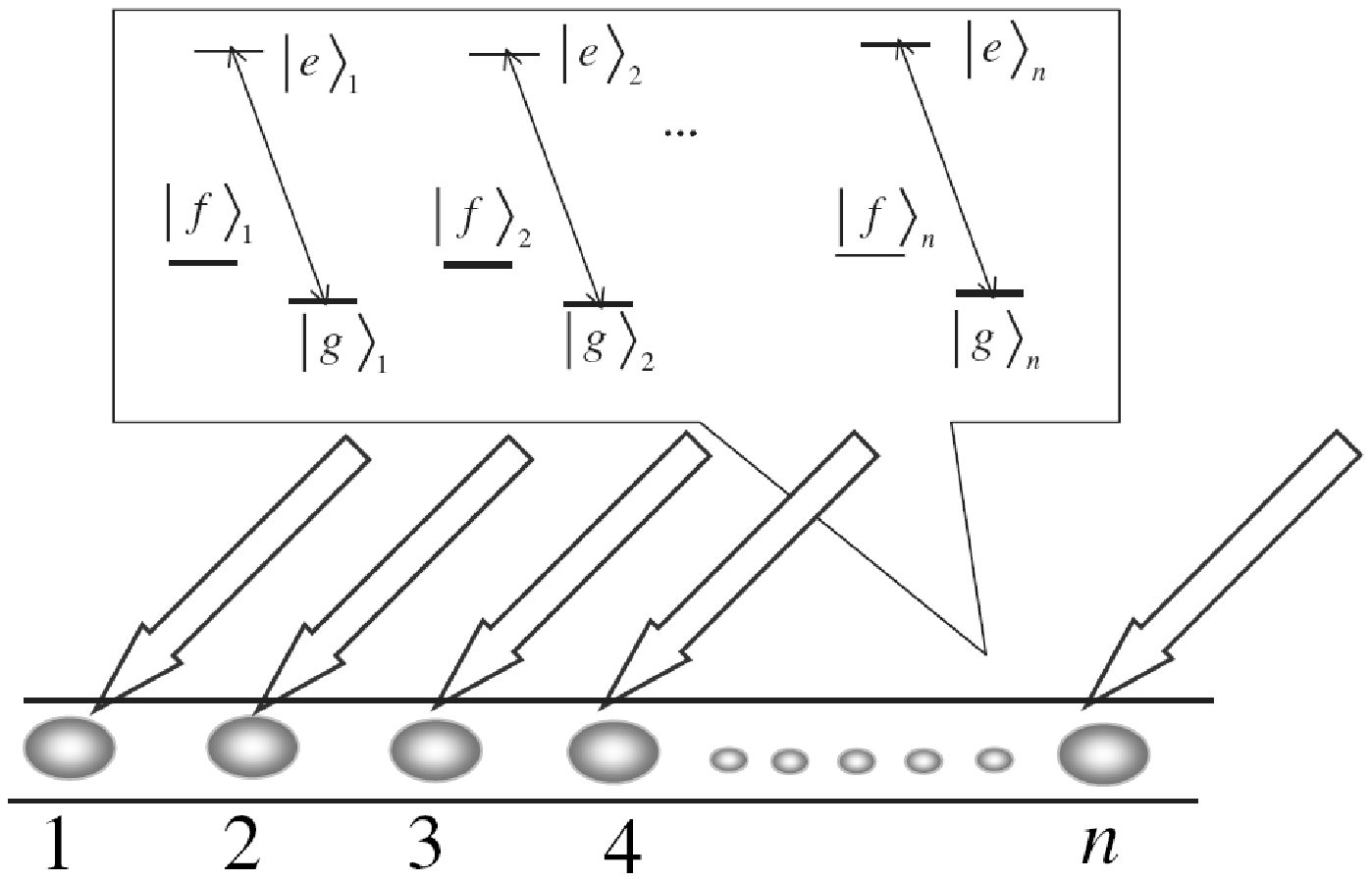}%
\end{center}
\end{figure}
%EndExpansion

\newpage

\newpage%
%TCIMACRO{\FRAME{ftbpF}{5.8937in}{4.4373in}{0in}{}{}{2.eps}%
%width 5.8937in;  height 4.4373in;  depth 0in;  original-width 5.834in;
%original-height 4.3863in;  cropleft "0";  croptop "1";  cropright "1";
%cropbottom "0";  filename '2.eps';file-properties "XNPEU";}}}%
%BeginExpansion
\begin{figure}
[ptb]
\begin{center}
\includegraphics[
height=4.4373in,
width=5.8937in
]%
{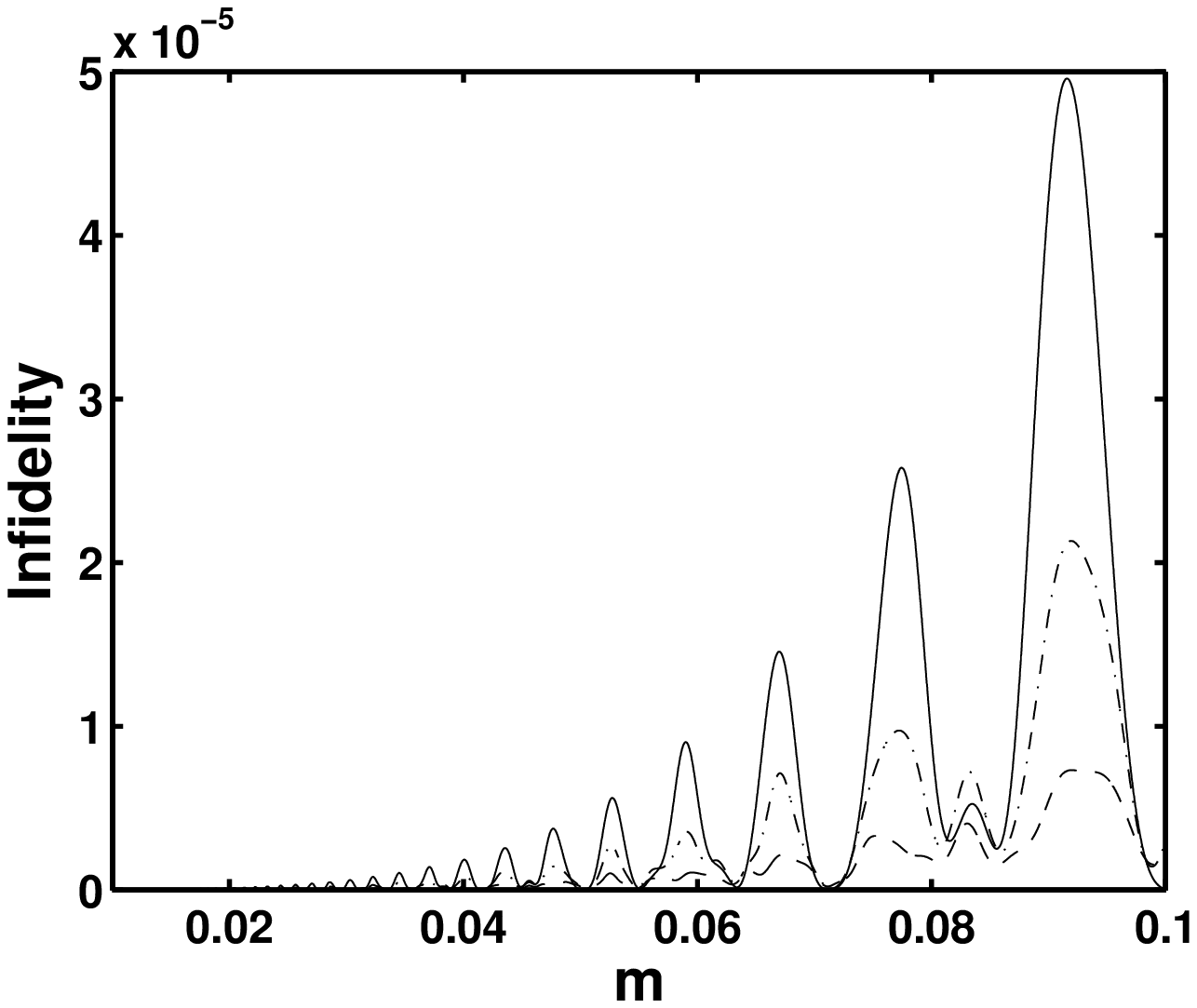}%
\end{center}
\end{figure}
%EndExpansion%
%TCIMACRO{\FRAME{ftbpF}{5.8937in}{4.4373in}{0in}{}{}{3.eps}%
%width 5.8937in;  height 4.4373in;  depth 0in;  original-width 5.834in;
%original-height 4.3863in;  cropleft "0";  croptop "1";  cropright "1";
%cropbottom "0";  filename '3.eps';file-properties "XNPEU";}}}%
%BeginExpansion
\begin{figure}
[ptbptb]
\begin{center}
\includegraphics[
height=4.4373in,
width=5.8937in
]%
{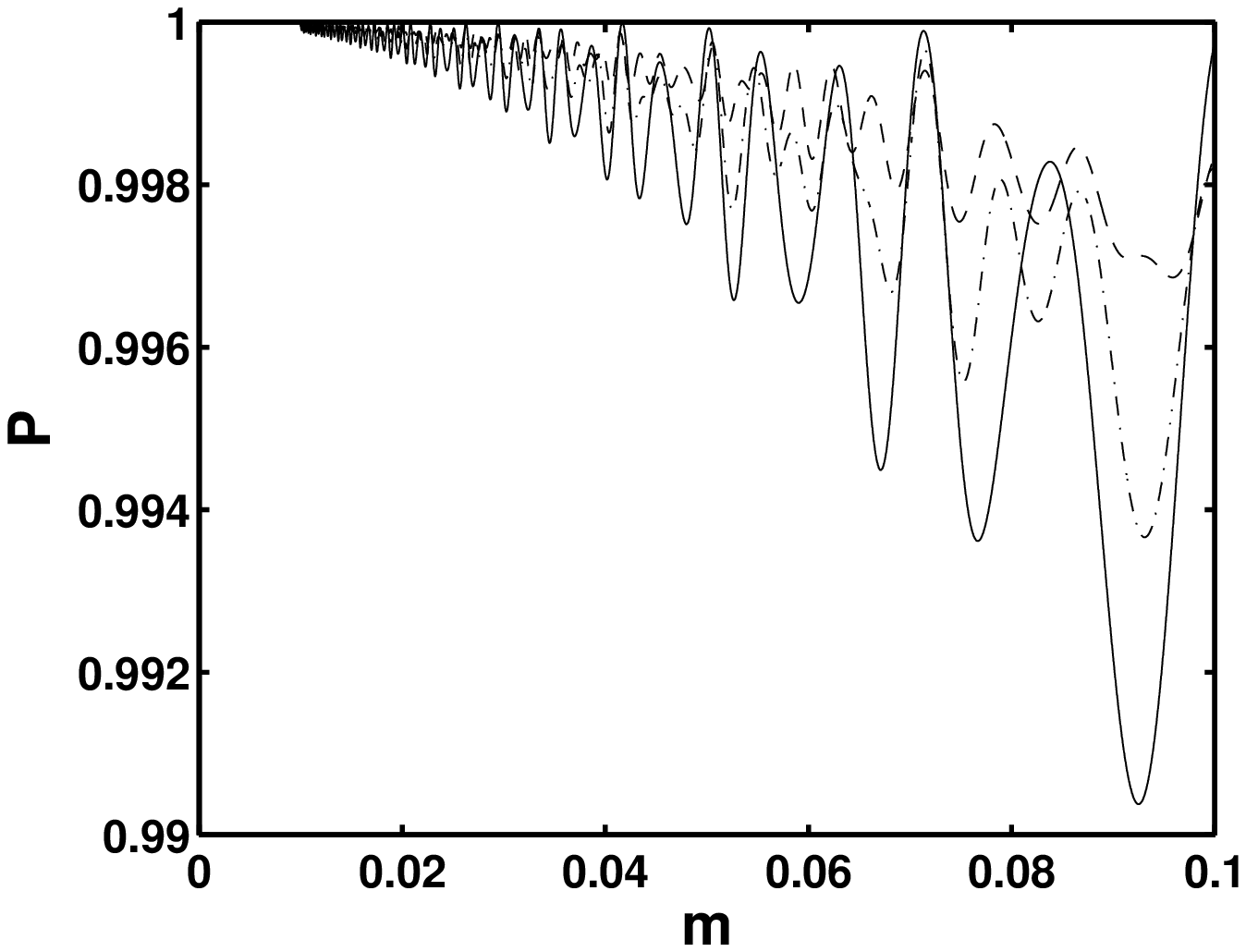}%
\end{center}
\end{figure}
%EndExpansion

\newpage%
%TCIMACRO{\FRAME{ftbpF}{7.8923in}{11.3602in}{0pt}{}{}{4.eps}%
%width 7.8923in;  height 11.3602in;  depth 0pt;  original-width 7.8646in;
%original-height 11.3334in;  cropleft "0";  croptop "1";  cropright "1";
%cropbottom "0";  filename '4.eps';file-properties "XNPEU";}}}%
%BeginExpansion
\begin{figure}
[ptb]
\begin{center}
\includegraphics[
height=11.3602in,
width=7.8923in
]%
{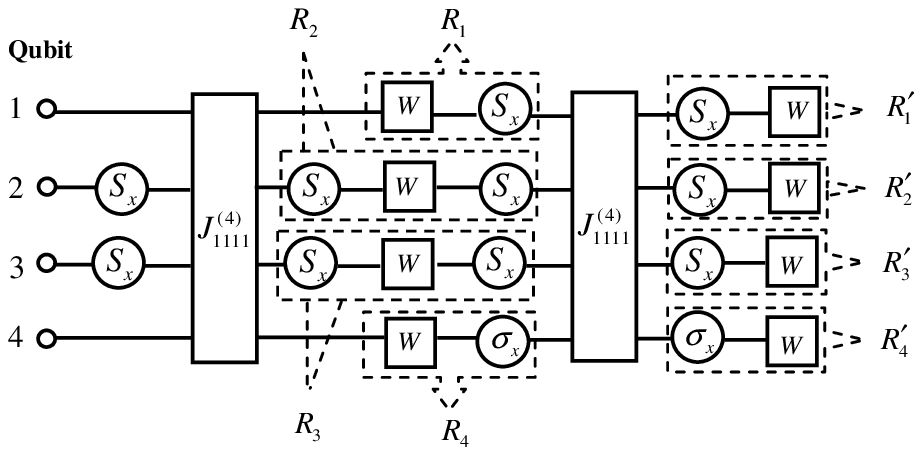}%
\end{center}
\end{figure}
%EndExpansion

\newpage%
%TCIMACRO{\FRAME{ftbpF}{7.8923in}{11.3602in}{0pt}{}{}{5.eps}%
%width 7.8923in;  height 11.3602in;  depth 0pt;  original-width 7.8646in;
%original-height 11.3334in;  cropleft "0";  croptop "1";  cropright "1";
%cropbottom "0";  filename '5.eps';file-properties "XNPEU";}}}%
%BeginExpansion
\begin{figure}
[ptb]
\begin{center}
\includegraphics[
height=11.3602in,
width=7.8923in
]%
{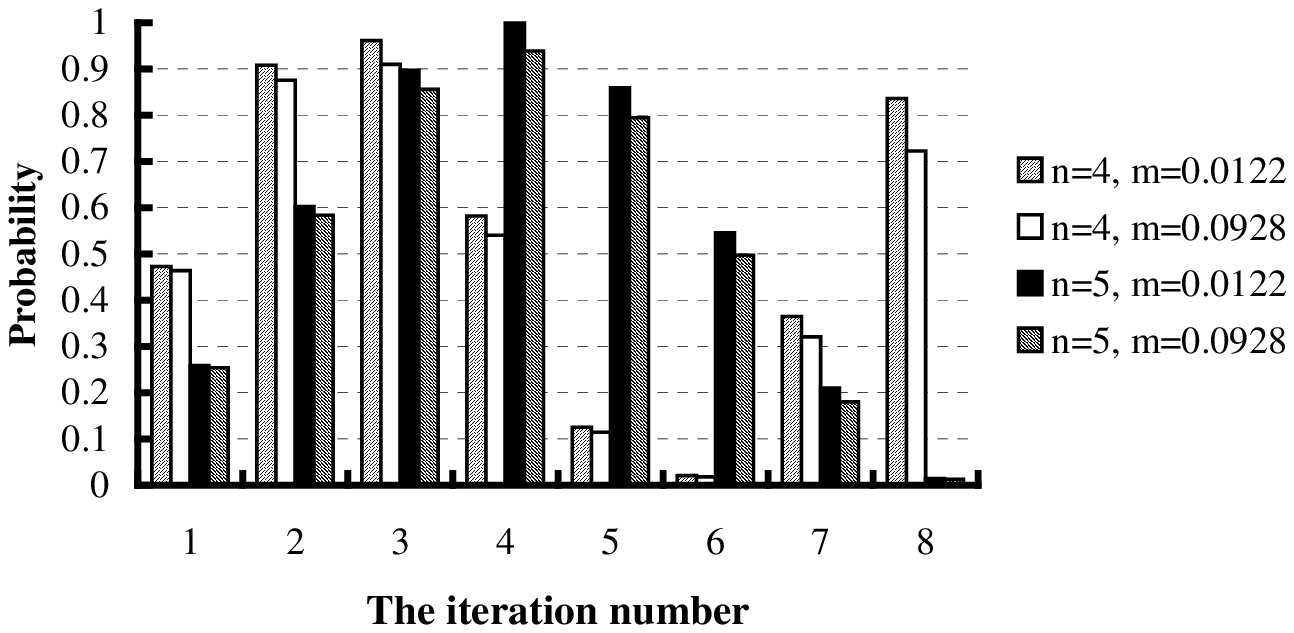}%
\end{center}
\end{figure}
%EndExpansion

\end{document}